\documentstyle[12pt,axodraw]{article}
\begin{document}
\vskip 0.2cm
\hfill{YITP-SB-04-64}
\vskip 0.2cm
\centerline{\large\bf {The difference between n-dimensional regularization and}}
\centerline{\large\bf {n-dimensional reduction in QCD. }}
\vskip 0.4cm
\centerline {\sc J. Smith 
\footnote{partially supported
by the National Science Foundation grant PHY-0354776.}}
\centerline{\it C.N. Yang Institute for Theoretical Physics,}
\centerline{\it State University of New York at Stony Brook,
New York 11794-3840, USA.}
\vskip 0.2cm
\centerline {\sc W.L. van Neerven}
\centerline{\it Instituut-Lorentz}
\centerline{\it University of Leiden,}
\centerline{\it PO Box 9506, 2300 RA Leiden,}
\centerline{\it The Netherlands.}
\vskip 0.2cm
\centerline{November 2004}
\vskip 0.2cm
\centerline{\bf Abstract}
\vskip 0.3cm
We discuss the difference between n-dimensional regularization and
n-dimensional reduction for processes in QCD which have an additional mass scale.
Examples are heavy flavour production in hadron-hadron collisions or on-shell photon-hadron 
collisions where the scale is represented by the mass $m$. Another example is
electroproduction of heavy flavours where we have two mass scales given by
$m$ and the virtuality of the photon $Q=\sqrt{-q^2}$. Finally we study the Drell-Yan process 
where the additional scale is represented by the virtuality $Q=\sqrt{q^2}$
of the vector boson ($\gamma^*, W, Z$). The difference between the two schemes is 
not accounted for by the usual oversubtractions. There are extra counter terms which
multiply the mass scale dependent parts of the Born cross sections. In the case of
the Drell-Yan process it turns out that the off-shell mass regularization agrees with
n-dimensional regularization. 

\vfill
\newpage
Here we discuss the consistency between n-dimensional regularization
and n-dimensional reduction for processes in Quantum Chromodynamics (QCD)
which have an additional mass scale. This paper is a continuation of earlier
work which dealt with jet physics in hadron-hadron collisions where no additional
mass scale is present \cite{kusi}, \cite{case}. The method of n-dimensional regularization
was originally introduced in \cite{hove} with one exception that
the number of degrees of freedom of the gluon is now taken to be n-2.
All numerators of virtual and radiative graphs are represented in n dimensions. Likewise 
the loop integrals and phase space integrals are evaluated in n dimensions.
For the gluon spin average one has the factor $1/(n-2)$. The method of
n-dimensional reduction was proposed in \cite{si} (see also \cite{cajo}).
Apart from the number of external dimensions which is 4 instead of n (see table \ref{tab1})
the numerators for virtual and radiative graphs are now presented for n equal to 4. 
However the loop integrals and phase space integrals are still evaluated in n dimensions.
This implies that the tensorial
reduction of the loop graphs and phase space integrals are still done
in n dimensions. Only traces and the usual Lorentz algebra are done in four dimensions.
The gluon spin average factor is now $1/2$.
We compare the two schemes in table \ref{tab1}.          
\begin{table}[hb]
\begin{center}
\begin{tabular}{|c|c|c|}\hline
& n-dim. regularization& n-dim. reduction\\ \hline
\hline
           &                &                             \\
number of internal dimensions &  n        &  n        \\[1mm]
number of external dimensions &  n  & 4  \\[2mm]
number of internal gluons &  n-2     &  2      \\[2mm]
number of external gluons & n-2  &  2 \\[2mm]
number of internal quarks & 2 &   2 \\[2mm]
number of external quarks & 2  &  2 \\[2mm] \hline
\end{tabular}
\end{center}
\caption{Definitions of the numbers of degees of freedom in the two 
regularization prescriptions.}
\label{tab1}
\end{table}
If we perform both regularization techniques the usual divergences which 
appear in next-to-leading order (NLO) calculations are
of the ultraviolet (UV), infrared (IR) and collinear (C) type and produce pole
terms of the type $1/(n-4)^k$. After cancelling the IR and the final state C divergences
by adding the results for the loop graphs to the squares of the radiative graphs we 
are left with the UV singularities
and the initial state C divergences. This is true for inclusive processes
only. Then we have to perform mass renormalization and coupling constant renormalization
to get rid of the UV divergences. In this paper we choose the on-mass-shell scheme
for mass renormalization in both regularization procedures, where
\begin{eqnarray}
\hat m=m\,\left [1+C_F\,\frac{\alpha_s}{4\pi}\,\left (\frac{6}{n-4}+3\,\gamma_E -3\,
\ln 4\pi -4 -3\,\ln \frac{\mu^2}{m^2}\right )\right ]\,.
\end{eqnarray}
Here $\hat m$ and $m$ denote the bare and renormalized mass respectively.
Coupling constant renormalization is achieved in n-dimensional regularization
in the ${\overline {\rm MS}}$ scheme by
\begin{eqnarray}
\hat \alpha_s&=&\alpha_s\,\left [1+\frac{\alpha_s}{4\pi}\,\beta_0\,
\left \{\frac{2}{n-4}+\gamma_E-\ln 4\,\pi\right \}\right ]\,,
\nonumber\\[2ex]
\beta_0&=&\frac{11}{3}\,C_A-\frac{4}{3}\,T_f\,n_f\,,
\end{eqnarray}
where $\hat \alpha_s$ and $\alpha_s$ denote the bare and renormalized coupling constant
respectively. The initial state C singularities
are removed via mass factorization. The latter is achieved by subtracting the Born
cross sections convoluted with kernels in
which the residues of the pole terms are given by the splitting functions $P_{ij}$
(for the normalization see (5.9) in \cite{rasm1}). Choosing the ${\overline {\rm MS}}$
scheme in n-dimensional regularization we have
\begin{eqnarray}
\Gamma_{ij}(x)&=&\delta_{ij}\,\delta(1-x)+\frac{\alpha_s}{4\pi}\,\left [
\frac{1}{2}\,P_{ij}(x)\,\left(\frac{2}{n-4}+\gamma_E-\ln 4\pi\right )\right ]\,.
\end{eqnarray}
\begin{figure}[t]
\begin{center}
\begin{picture}(300,100)(0,0)

\Gluon(10,10)(90,10){3}{8}
\Gluon(50,10)(50,50){3}{4}
\Gluon(10,90)(50,90){3}{4}
\ArrowLine(50,50)(90,50)
\ArrowLine(50,90)(50,50)
\ArrowLine(90,90)(50,90)

\DashArrowLine(210,10)(250,10){3}
\DashArrowLine(250,10)(290,10){3}
\Gluon(250,10)(250,50){3}{4}
\Gluon(210,90)(250,90){3}{4}
\ArrowLine(250,50)(290,50)
\ArrowLine(250,90)(250,50)
\ArrowLine(290,90)(250,90)

\Text(60,30)[t]{$t$}
\Text(260,30)[t]{$t$}
\Text(0,100)[t]{$k_2$}
\Text(0,15)[t]{$k_1$}
\Text(100,95)[t]{$p_2$}
\Text(100,55)[t]{$p_1$}

\Text(200,100)[t]{$k_2$}
\Text(200,15)[t]{$k_1$}
\Text(300,95)[t]{$p_2$}
\Text(300,55)[t]{$p_1$}

\end{picture}
\caption[]{ t-channel graphs for the heavy flavour production processes $g+g \rightarrow
Q+\bar Q +g$ and $g+q \rightarrow Q+\bar Q +q$.}
\label{fig1}
\end{center}
\end{figure}
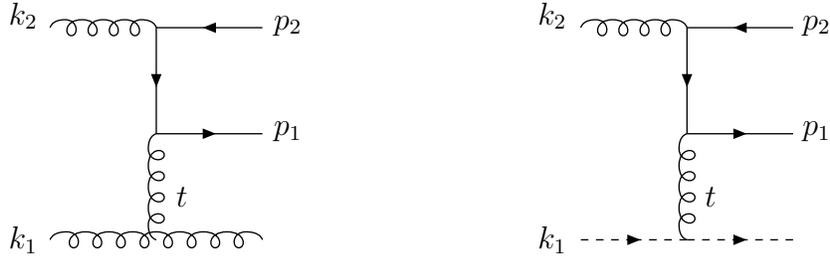
It is clear that both regularization procedures
lead to finite results which are however different. These differences can be accounted for
by performing an additional finite coupling constant renormalization and a
finite mass factorization in the case of n-dimensional reduction. Here we use
\begin{eqnarray}
\hat \alpha_s&=&\alpha_s\,\left [1+\frac{\alpha_s}{4\pi}\,
\left \{\beta_0\left (\frac{2}{n-4}+\gamma_E-\ln 4\,\pi\right )
+C_A\,\frac{1}{3}\right \}\right ]\,,
\end{eqnarray}
and
\begin{eqnarray}
\Gamma_{ij}(x)=\delta_{ij}\,\delta(1-x)+\frac{\alpha_s}{4\pi}\,\left [
\frac{1}{2}\,P_{ij}(x)\,\left(\frac{2}{n-4}+\gamma_E-\ln 4\pi\right )+Z_{ij}(x)\right ]\,,
\end{eqnarray}
with \cite{blra} \footnote{ In \cite{blra} the $-4\,x+4\,x^2$ must be put in the upper 
righthand corner}
\begin{eqnarray}
Z=\left (\begin{array}{cc}
Z_{qq}&Z_{qg}\\
Z_{gq}&Z_{gg}
\end{array}
\right )=
\left (\begin{array}{cc}
C_F\,[-2+2\,x+\delta(1-x)]&T_f\,[ -4\,x+4\,x^2]\\
C_F\,[-2\,x]& C_A\,[\delta(1-x)/3]
\end{array}
\right )\,.
\end{eqnarray}
For $SU(N)$ we have $C_A=N$, $C_F=(N^2-1)/2\,N$ and $T_f=1/2$. In QCD we have $N=3$. 

In this paper we shall concentrate on the radiative graphs and reserve some comments on
the loop graphs for the end. This implies that we will limit our discussions to
the regular parts of the kernels $\Gamma_{ij}$ and postpone the treatment of the 
singular parts represented by the $\delta(1-x)$ terms to later on.
With the above finite coupling constant renormalization
and finite mass factorization the jet cross sections in hadron-hadron collisions
\cite{kusi}, \cite{case} could be brought into agreement with each other.
However for processes which have an additional mass scale this was not successful \cite{beku}.
Here we have additional terms which however only multiply the mass dependent parts of the 
Born cross sections. Therefore these are not finite mass factorizations because
they would involve the whole Born cross sections. These additional terms are given by
\begin{eqnarray}
K_i=C_i\,\frac{\alpha_s}{4\pi}\,\left [-4\,\frac{1-x}{x}\right ] \,, \qquad
C_q=C_F\,, \quad C_g=C_A \,,
\end{eqnarray}
and are universal. Careful examination shows that they only occur for unpolarized 
processes which have a gluon exchange in a t-channel or a u-channel
graph, see for instance the diagrams in Fig. \ref{fig1}. 
The terms are of the type $m^2/t^2$ or $m^2/u^2$
where $m$ is the additional mass scale.
For polarized processes this phenomenon does not occur because these terms do not
exist. This can be inferred from Eq. 7 because of the 
term $1/x$ which is characteristic for unpolarized processes which have a gluon exchange
in the t-channel or u-channel. We now examine specific reactions.

Let us start with heavy flavour production in hadron-hadron collisions. In \cite{beku}
the cross sections were calculated in both regularization schemes. For the $gg$
channel the Born cross section can be written as follows (see Eqs. (2.5)-(2.11) in
\cite{beku})
\begin{eqnarray}
&& g(k_1)+g(k_2) \rightarrow Q(p_1)+\bar Q(p_2)\,,
\nonumber\\[2ex]
&&s=(k_1+k_2)^2 \,,\qquad t_1=(k_2-p_2)^2-m^2 \,,\qquad u_1=(k_1-p_2)^2-m^2\,,
\nonumber\\[2ex]
&& s^2\,\frac{d^2\sigma^{(0)}_{gg}}{dt_1~du_1}=
s^2\,\frac{d^2\sigma^{(0)}_{gg,O}}{dt_1~du_1}
+s^2\,\frac{d^2\sigma^{(0)}_{gg,K}}{dt_1~du_1}\,,
\nonumber\\[2ex]
&&s^2\,\frac{d^2\sigma^{(0)}_{gg,O}}{dt_1~du_1}=\pi\,\alpha_s^2\,
\frac{N}{2\,(N^2-1)}\,\left [\frac{t_1^2+u_1^2}{s^2}\right ]\,B_{QED}\,\delta(s+t_1+u_1)\,,
\nonumber\\[2ex]
&&s^2\,\frac{d^2\sigma^{(0)}_{gg,K}}{dt_1~du_1}=-\pi\,\alpha_s^2\,
\frac{1}{2\,N\,(N^2-1)}\,B_{QED}\,\delta(s+t_1+u_1)\,,
\nonumber\\[2ex]
&&B_{QED}=\frac{t_1}{u_1}+\frac{u_1}{t_1}+\frac{4\,m^2\,s}{t_1\,u_1}\,
\left (1-\frac{m^2\,s}{t_1\,u_1} \right )\,.
\end{eqnarray}
We encounter for the first time differences between n-dimensional regularization 
and n-dimensional reduction in the (NLO) $gg$ cross sections 
(see Eq. (6.16) and (6.17) in \cite{beku}). They are represented by 
the terms $K_g$ convoluted with the mass dependent parts of the Born cross sections indicated by 
the subscript $m$
\begin{eqnarray}
s^2\,\frac{d^2\sigma^{(1)}_{gg,i}}{dt_1~du_1}\Bigg |_{reg}=
s^2\,\frac{d^2\sigma^{(1)}_{gg,i}}{dt_1~du_1}\Bigg |_{red}+
2\,K_g\otimes\,
s^2\,\frac{d^2\sigma^{(0)}_{gg,i}}{dt_1~du_1}\Bigg |_m\qquad \,, i=O,K\,,
\end{eqnarray}
where the symbol $\otimes$ denotes the convolution integral.
The $gg$ Born cross sections can also be written in a different way, namely 
\begin{eqnarray}
s^2\,\frac{d^2\sigma^{(0)}_{gg,F}}{dt_1~du_1}&=&\pi\,\alpha_s^2\,
\frac{C_F}{N^2-1}\,B_{QED}\,\delta(s+t_1+u_1)\,,
\nonumber\\[2ex]
s^2\,\frac{d^2\sigma^{(0)}_{gg,A}}{dt_1~du_1}&=&-\pi\,\alpha_s^2\,
\frac{C_A}{N^2-1}\,\left (\frac{t_1u_1}{s^2}\right )\,B_{QED}\,\delta(s+t_1+u_1)\,.
\end{eqnarray}
Moreover we have the Born cross section for the $q\bar q\rightarrow Q\bar Q$ reaction
\begin{eqnarray}
s^2\,\frac{d^2\sigma^{(0)}_{q\bar q}}{dt_1~du_1}&=&\pi\,\alpha_s^2\,
\frac{C_F}{N}\,A_{QED}\,\delta(s+t_1+u_1)\,,
\nonumber\\[2ex]
A_{QED}&=&\frac{t_1^2+u_1^2}{s^2}+\frac{2\,m^2}{s}\,.
\end{eqnarray}
In this way Eq. (4.23) in \cite{bene} can be written as
\begin{eqnarray}
&&s^2\,\frac{d^2\sigma^{(1)}_{g\bar q,F}}{dt_1~du_1}\Bigg |_{reg}=
s^2\,\frac{d^2\sigma^{(1)}_{g\bar q,F}}{dt_1~du_1}\Bigg |_{red}
+Z_{gq}\otimes s^2\,\frac{d^2\sigma^{(0)}_{gg,F}}{dt_1~du_1}
\nonumber\\[2ex]
&&+Z_{qg}\otimes s^2\,\frac{d^2\sigma^{(0)}_{q\bar q}}{dt_1~du_1}
+K_q\otimes\,
s^2\,\frac{d^2\sigma^{(0)}_{gg,F}}{dt_1~du_1}\Bigg |_m\,,
\end{eqnarray}
and Eq. (4.24) becomes equal to
\begin{eqnarray}
&&s^2\,\frac{d^2\sigma^{(1)}_{g\bar q,A}}{dt_1~du_1}\Bigg |_{reg}=
s^2\,\frac{d^2\sigma^{(1)}_{g\bar q,A}}{dt_1~du_1} \Bigg |_{red}
+Z_{gq}\otimes s^2\,\frac{d^2\sigma^{(0)}_{gg,A}}{dt_1~du_1}
\nonumber\\[2ex]
&&+K_q\otimes\,s^2\,\frac{d^2\sigma^{(0)}_{gg,A}}{dt_1~du_1} \Bigg |_m\,.
\end{eqnarray}
Both cross sections involve extra terms which are proportional to
$K_q$ convoluted with the mass dependent parts of the cross sections.
Finally the $q\bar q$ cross section behaves in a normal way and Eq. (4.8) in \cite{bene}
becomes
\begin{eqnarray}
s^2\,\frac{d^2\sigma^{(1)}_{q\bar q,F}}{dt_1~du_1}_{reg}=
s^2\,\frac{d^2\sigma^{(1)}_{q\bar q,F}}{dt_1~du_1}_{red}
+2\,Z_{qq}\otimes s^2\,\frac{d^2\sigma^{(0)}_{q\bar q}}{dt_1~du_1}\,.
\end{eqnarray}

The next process is electroproduction of heavy flavours. Here two
mass scales are involved i.e. the heavy flavour mass $m$ and the virtuality of the 
off-shell photon $Q^2=-q^2$. The Born cross sections for the transverse (G) and
longitudinal (L) parts are (see Eqs. (2.14) and (2.15) in \cite{lari})
\begin{eqnarray}
&&\gamma^*(q) +g(k_1) \rightarrow Q(p_1)+\bar Q(p_2)\,, \qquad
s=(q+k_1)^2=s'+q^2 \,,
\nonumber\\[2ex]
&& t_1=(k_1-p_2)^2-m^2\,, \qquad u_1=(q-p_2)^2-m^2=u_1'+q^2\,,
\nonumber\\[2ex]
&&s'^2\,\frac{d^2\sigma^{(0)}_{i,g}}{dt_1~du_1}=\pi\,e_H^2\,\alpha\,\alpha_s
\,a_i\,B_{i,QED}\,\delta(s'+t_1+u_1) \,, \qquad i=G,L\,,
\nonumber\\[2ex]
&& a_G=1 \,, \qquad a_L=2\,,
\nonumber\\[2ex]
&&B_{G,QED}=\frac{t_1}{u_1}+\frac{u_1}{t_1}+\frac{4\,m^2\,s'}{t_1\,u_1}\,
\left (1-\frac{m^2\,s'}{t_1\,u_1} \right )+\frac{2\,s'\,q^2}{t_1\,u_1}
+\frac{2\,q^4}{t_1\,u_1}
\nonumber\\[2ex]
&&+\frac{2\,m^2\,q^2}{t_1\,u_1}\,\left (2-\frac{s^{'2}}{t_1\,u_1}
\right )\,,
\nonumber\\[2ex]
&& B_{L,QED}=-\frac{4\,q^2}{s'}\,\left(1-\frac{q^2}{s'}-\frac{m^2\,s'}{t_1\,u_1}\right )\,,
\qquad q^2=-Q^2\,.
\end{eqnarray}
The differences between n-dimensional regularization and n-dimensional reduction 
are visible in the NLO off-shell photon-gluon fusion processes. 
Eqs. (4.7) and (4.8) in \cite{lari} are equal to
\begin{eqnarray}
s^2\,\frac{d^2\sigma^{(1)}_{i,g}}{dt_1~du_1}\Bigg |_{reg}=
s^2\,\frac{d^2\sigma^{(1)}_{i,g}}{dt_1~du_1}\Bigg |_{red}+
K_g\otimes\,
s^2\,\frac{d^2\sigma^{(0)}_{i,g}}{dt_1~du_1}\Bigg |_{m,Q}\,, \quad i=G,L\,,
\end{eqnarray}
where the second terms of the above equation contains all pieces proportional
to $m^2$ and $q^2$ in the Born cross sections in Eq. (15).
For the Bethe-Heitler process ($A_1$) in off-shell photon-quark scattering we see the same
phenomenon. Eq. (4.11) in \cite{lari} becomes 
\begin{eqnarray}
&&s^2\,\frac{d^2\sigma^{(1)}_{i,q,A_1}}{dt_1~du_1} \Bigg |_{reg}=
s^2\,\frac{d^2\sigma^{(1)}_{i,q,A_1}}{dt_1~du_1}\Bigg |_{red}
+Z_{gq}\otimes s^2\,\frac{d^2\sigma^{(0)}_{i,g}}{dt_1~du_1}
\nonumber\\[2ex]
&&+K_q\otimes\,
s^2\,\frac{d^2\sigma^{(0)}_{i,g}}{dt_1~du_1}\Bigg |_{m,Q}\,, \quad i=G,L\,.
\end{eqnarray}
For $i=G$ we get the same in the case of on-shell photon-hadron production ($q^2=0$)
\cite{smne} except that
now also the Compton process ($A_2$) gets a collinear divergence. The difference between both
regularizations in Eq. (4.17) of \cite{smne} becomes
\begin{eqnarray}
s^2\,\frac{d^2\sigma^{(1)}_{\gamma q,A_2}}{dt_1~du_1}\Bigg |_{reg}=
s^2\,\frac{d^2\sigma^{(1)}_{\gamma q,A_2}}{dt_1~du_1}\Bigg |_{red}
+Z_{qg}\otimes s^2\,\frac{d^2\sigma^{(0)}_{qq}}{dt_1~du_1}\,,
\end{eqnarray}
which is of the usual form.

Finally we turn our attention to the Drell-Yan process. We look at the differential
distributions of the vector boson with momentum $q$. The cross sections
have been calculated in n-dimensional regularization in \cite{arre}, \cite{gopa}.
We have also calculated them using n-dimensional reduction. The Born processes
and Born cross sections are given by
\begin{eqnarray}
&&q(p_1)+\bar q(p_2)\rightarrow \gamma^*(q) +g(k)\,, \qquad q(p_1)+ g(p_2)\rightarrow  
\gamma^*(q) + q(k)\,,
\nonumber\\[2ex]
&&s=(p_1+p_2)^2 \,, \qquad t=(p_1-q)^2 \,, \qquad u=(p_2-q)^2\,,\qquad q^2=Q^2\,,
\nonumber\\[2ex]
&&s^2\,\frac{d^2W^{(0)}_{q\bar q}}{dt~du}=\alpha_s\,\frac{C_F}{N}\,
\left [ \frac{4\,s\,Q^2+2\,t^2+2\,u^2}{t\,u}\right ]\,\delta(s+t+u-Q^2)\,,
\nonumber\\[2ex]
&&s^2\,\frac{d^2W^{(0)}_{qg}}{dt~du}=\alpha_s\,\frac{T_f}{N}\,
\left [ - \frac{4\,Q^2\,(Q^2-s-u)+2\,s^2+2\,u^2}{s\,u}\right ]
\, \delta(s+t+u-Q^2)\,.
\nonumber\\
\end{eqnarray}
The NLO $q\bar q$ process involves no problem. We find
\begin{eqnarray}
s^2\,\frac{d^2W^{(1)}_{q\bar q}}{dt~du} \Bigg |_{reg}=
s^2\,\frac{d^2W^{(1)}_{q\bar q}}{dt~du}\Bigg |_{red}
+2\,Z_{qq}\otimes s^2\,\frac{d^2W^{(0)}_{q\bar q}}{dt~du}\,.
\end{eqnarray}
However for the NLO $qg$ and $qq$ subprocesses differences
between n-dimensional regularization and n-dimensional reduction appear again
and equal the mass (here $Q^2$) dependent part of the Born cross sections
convoluted with either $K_g$ ($qg$) or $K_q$ ($qq$). 
\begin{eqnarray}
s^2\,\frac{d^2W^{(1)}_{qg}}{dt~du} \Bigg |_{reg}&=&
s^2\,\frac{d^2W^{(1)}_{qg}}{dt~du} \Bigg |_{red}
+Z_{qq}\otimes s^2\,\frac{d^2W^{(0)}_{qg}}{dt~du}
\nonumber\\[2ex]
&&+Z_{qg}\otimes s^2\,\frac{d^2W^{(0)}_{q\bar q}}{dt~du}
+K_g\otimes\,s^2\,\frac{d^2W^{(0)}_{qg}}{dt~du}\Bigg |_Q \,,
\\[2ex]
\nonumber\\[2ex]
s^2\,\frac{d^2W^{(1)}_{qq}}{dt~du}\Bigg |_{reg}&=&
s^2\,\frac{d^2W^{(1)}_{qq}}{dt~du}\Bigg |_{red}
+2\,Z_{gq}\otimes s^2\,\frac{d^2W^{(0)}_{qg}}{dt~du}
\nonumber\\[2ex]
&& +2\,K_q\otimes\,s^2\,\frac{d^2W^{(0)}_{qg}}{dt~du}\Bigg |_Q \,.
\end{eqnarray}
Finally the NLO $gg$ subprocess behaves like the $q\bar q$ subprocess and does 
need this extra term, so
\begin{eqnarray}
s^2\,\frac{d^2W^{(1)}_{gg}}{dt~du} \Bigg |_{reg}=
s^2\,\frac{d^2W^{(1)}_{gg}}{dt~du} \Bigg |_{red}
+2\,Z_{qg}\otimes s^2\,\frac{d^2W^{(0)}_{qg}}{dt~du}\,.
\end{eqnarray}
In all the above reactions we observe that when the mass dependent part of the cross section
appears convoluted with $K_k$ ($k=q,g$) we also encounter the exchange of a gluon
in t or u channel graphs. 

To decide which regularization prescription is correct we try
out another regularization technique. Here we choose the off-shell technique 
\cite{alel}, \cite{haka}, \cite{hune1} which
is defined so that all external particles are taken off-shell $p_i^2<0$. The intrinsic
particle masses are equal to zero and the collinear divergences
appear as $\ln (-Q^2/p^2)$. The kernels $\Gamma_{ij}$ become equal to
the operator matrix elements where the external legs are put off-shell.
In this case the regular part of $\Gamma_{ij}$ in the ${\overline {\rm MS}}$ scheme becomes
\begin{eqnarray}
\Gamma_{ij}(x)=\delta_{ij}\,\delta(1-x)+\frac{\alpha_s}{4\pi}\,\left [
\frac{1}{2}\,P_{ij}(x)\,\ln \left (\frac{\mu^2}{-x\,(1-x)\,p^2}\right )+Z_{ij}(x)\right ]\,,
\end{eqnarray}
with the finite renormalization $Z$ equal to
\begin{eqnarray}  
Z(x)=
\left (\begin{array}{cc}
C_F\,[-4+2\,x]&T_f\,[ -2 -4\,x\,(1-x)]\\
C_F\,[(-4+2\,x-2\,x^2)/x]& C_A\,[(5\,x-4)/x]
\end{array}
\right )\,.
\end{eqnarray}
We omit the $\delta(1-x)$ terms in $Z_{ij}$ because they concern the soft-plus-virtual
gluon contributions. These terms are very complicated in the off-shell approach, see \cite{hune2}.
Substituting $\Gamma_{ij}$ in the above equations we observe that $K_i=0$, in other
words we get the same as n-dimensional regularization. Apparently the $n-4$ terms
appearing in the numerator by use of n-dimensional regularization, which are multiplied by
pole terms $1/(n-4)$, mimic analogous terms in the numerator which are proportional to $p^2$ 
in the case of off-shell regularization and are multiplied by $1/p^2$. The latter 
terms arise in those parts of the cross sections which are proportional to $p^2/t^2$
or $p^2/u^2$. Therefore one cannot omit these terms.
In n-dimensional reduction the $n-4$ terms are not present
and $p^2=0$ is put at the beginning. This leads us to the conclusion
that for QCD processes with an additional mass scale n-dimensional reduction is
wrong unless one wants to add an additional mass factorization which 
however is not proportional to the whole Born cross section. 

Finally we have also studied
the soft-plus-virtual gluon contributions in the n-dimensional regularization
and n-dimensional reduction methods. Since in the loop graphs UV divergences also appear
we only get consistency between both regularizations if we choose an ${\cal }N=1$
supersymmmetry where the quarks are now Majorana fermions in the adjoint representation.
Therefore $C_A=C_F=n_f=N$ for $SU(N)$. For the Drell-Yan $q\bar q$ process we get consistency
provided we implement the finite coupling constant renormalization in Eq. (4) and
the finite mass factorization in Eq. (6) for the $\delta(1-x)$
terms. However for the $qg$ process we get an inconsistency unless we put a factor
of 3/2 instead of a one in the coefficient of the term containing the $\delta(1-x)$ 
function in $Z_{qq}$ of Eq. (6). This is
in disagreement with what we found for the Drell-Yan $q\bar q$ process and with the
jet processes in \cite{kusi} and \cite{case}.

In conclusion we find a disagreement in the radiative part of
the NLO cross sections between 
n-dimensional regularization and n-dimensional reduction for processes which 
involve an additional mass scale. However the off-shell regularization method indicates 
that n-dimensional regularization yields the correct answer.
%

\end{document}